\documentstyle[aps,preprint,epsf]{revtex}
\begin{document}
\draft

\title{On the quantum probability flux through surfaces}

\author{M.~Daumer, D.~D\"urr}
\address{Fakult\"at f\"ur Mathematik, Universit\"at M\"unchen,
      Theresienstr. 39, 80333 M\"unchen, Germany}
\author{S.~Goldstein}
\address{Department of Mathematics, Rutgers University, New Brunswick,
      New Jersey 08903, USA}
\author{ N.~Zangh\`{\i}}
\address{Istituto di Fisica, Universit\`a di Genova, INFN, 
Via Dodecaneso 33, 16146 Genova, Italy}
 
\date{January 23, 1997}
\maketitle

\begin{abstract}

We remark that the often ignored quantum probability
current is fundamental for a genuine understanding of scattering phenomena
and, in particular, for the statistics of the time and position of the
first exit of a quantum particle from a given region, which may be simply
expressed in terms of the current. This simple formula for these statistics
does not appear as such in the literature. It is proposed that the formula,
 which is very different from the usual quantum mechanical measurement
formulas, be verified experimentally.
A full understanding of the quantum current and the associated formula
is provided by Bohmian mechanics.

\end{abstract}
 



\bigskip

\section{Introduction}
In  Born's interpretation of the wave function $\psi_t$ at time $t$
of a single particle of mass $m$, $\rho_t({\bf x})=|\psi_t({\bf x})|^2$ is the
probability density for finding the particle at ${\bf x}$ at that time.
The consistency of this interpretation is ensured by the continuity
equation
$$
\frac{\partial \rho_t}{\partial t} + {\rm div \/ } \cdot {\bf j}^{\psi_t} = 0, 
$$
where ${\bf j}^{\psi_t} = \frac{1}{m}{\rm Im \/ }\psi_t^* \nabla \psi_t$ is the
quantum current ($\hbar = 1$).

The quantum current is usually not considered to be of any operational
significance (see however \cite{Aharonov}). It is not related to any
standard quantum mechanical measurement in the way, for example, that the
density $\rho$, as the spectral measure of the position operator, gives the
statistics for a position measurement. Nonetheless, it is hard to resist
the suggestion that the quantum current integrated over a surface gives the 
probability that the particle crosses that surface,
i.e., that
\begin{equation} \label{JDOTDSIGMA} {\bf j}^{\psi_t} \cdot d{\bf S} dt 
\end{equation} 
is the probability  that a particle crosses the
surface element $d{\bf S}$ in the time $dt$. However, this suggestion must
be taken ``cum grano salis'' since ${\bf j}^{\psi_t} \cdot
d{\bf S} dt$ may be somewhere negative, in which case it cannot be a
probability. But before discussing the situations where ${\bf j}^{\psi_t}
\cdot d{\bf S} dt$ can be negative we want to consider first a regime for
which we can expect this quantity to be positive, so that its meaning could
in fact be the crossing probability, namely, the regime described by
scattering theory.

\section{Standard Scattering Theory}
In
textbooks on quantum mechanics the principal objects of interest for
scattering phenomena are nonnormalized stationary solutions of the 
Schr\"odinger equation with the asymptotic behavior
$$
\psi({\bf x}) \stackrel{x\to\infty}{\sim} e^{i{\bf p}_{in}\cdot{\bf x}} + f(\theta,\phi) \frac{e^{ipx}}{x},  
$$
where $e^{i{\bf p}_{in}\cdot{\bf x}}$ represents  an  incoming  
wave, $p=|{\bf p}_{in}|$, and
$f(\theta,\phi) \frac{e^{ipx}}{x}$ is the scattered  wave with
angular dependent amplitude. $f(\theta,\phi)$ gives  the probability for deflection of the
particle in the direction specified by $\theta,\phi$ by the well-known formula
for the differential cross section
\begin{equation}
\label{BORNFORM}
d\sigma = |f(\theta,\phi)|^2 \sin \theta d\theta d\phi
\end{equation}
This representation of a scattering process is, however, not entirely  convincing since
Born's rule is not directly applicable to non-normalizable wave functions.
More important, this picture is entirely time-independent whereas the physical scattering
event  is certainly a process in space and time.
Indeed, according to some experts, the arguments leading to
the formula (\ref{BORNFORM}) for the cross section 
``wouldn't convince an educated first grader'' (\cite{Simon}, p.~97).

It is widely accepted that the stationary treatment is
justified by an analysis of wave packets evolving with time.  Using a
normalized wave packet $\psi_t({\bf x}) = e^{-iHt}\psi({\bf x})$ one
immediately obtains by Born's rule the probabilities for position
measurements. But what are the relevant probabilities in a scattering
experiment? In mathematical physics (e.g.\ \cite{ReedSimon3}, p.~356, and
\cite{Enss}) an answer to this is provided by Dollard's
scattering-into-cones theorem\ \cite{Dollard}:
$$
\lim_{t \to \infty} \int_C d^3x |\psi_t({\bf x})|^2 = 
\int_C d^3p |\widehat{\Omega_-^\dagger\psi}({\bf p})|^2 . 
$$
This connects the asymptotic probability of finding the particle in
some cone $C$ with the probability of finding its asymptotic momentum 
${\bf p}$ in that cone, where
$\Omega_- = {\mbox{\rm s-}} \lim_{t\to\infty} e^{iHt}e^{-iH_0t}$ is the  wave
operator (``${\mbox{\rm s-}} \lim$'' denotes the strong limit), $H=H_0+V$, with
$H_0 = -\frac{1}{2m}\nabla^2$,  
 and $\hat{}$ denotes the
Fourier transform. It is generally believed that the left hand side of the 
scattering-into-cones theorem\ is exactly what
the scattering experiment measures, as if the fundamental cross section 
associated with the solid angle $\Sigma$ (to be identified with a subset of the unit sphere)
were
$$
\sigma_{{\mbox{\rm \footnotesize cone}}} (\Sigma) := \lim_{t \to \infty} 
\int_{C_\Sigma} d^3x |\psi_t({\bf x})|^2,
$$
where $C_\Sigma$ is the cone with apex at the origin subtended by $\Sigma$
(see Fig.~1).
To connect this with (\ref{BORNFORM}),
which is independent
of the details of the initial wave function, one may invoke the right hand side
of the scattering-into-cones theorem\ to 
recover the usual formula with additional assumptions on the initial wave packet 
(see \cite{ReedSimon3} p.\  356 for a discussion of this.)

So far the mathematics. But back to physics. The left hand side of the
scattering-into-cones theorem\ is the probability that at some large but fixed time, when the
position of the particle is measured, the particle is found in the cone
$C$. But does one actually measure in a scattering experiment in what cone
the particle happens to be found at some large but {\it fixed} time?  Is it not rather
the case that one of a collection of distant detectors surrounding the
scattering center fires at some {\it random} time, a time that is not
chosen by the experimenter? And isn't that random time simply the time at
which, roughly speaking, the particle crosses the detector
surface subtended by the cone?  

This suggests that the relevant quantity for the scattering experiment
should be the quantum current.  If the detectors are sufficiently distant from
the scattering center the current will typically be outgoing and
(\ref{JDOTDSIGMA}) will be positive. We obtain as the probability that the
particle has crossed some distant surface during some time interval the
integral of (\ref{JDOTDSIGMA}) over that time interval and that
 surface.  The integrated current thus provides us with a physical definition
(see also \cite{Newton}, p.\ 164) of the cross section: 
\begin{equation}
\label{SIGMAFLUX} \sigma_{{\mbox{\rm \footnotesize flux}}} (\Sigma) := 
\lim_{R \to \infty} \int_0^\infty dt
\int_{R\Sigma} {\bf j}^{\psi_t} \cdot d{\bf S}, \end{equation} 
where $R\Sigma$ is the intersection of the cone $C_\Sigma$ with the sphere
of radius $R$ (see Fig.~1).
 As before, one would like to connect this with the usual
formulas and hence we need
 the counterpart of the scattering-into-cones theorem---the 
flux-across-surfaces theorem---which provides us with a formula 
for $\sigma_{{\mbox{\rm \footnotesize  flux}}}$:
\begin{equation}
\label{FAST}
\lim_{R \to \infty} \int_0^\infty dt \int_{R\Sigma} {\bf j}^{\psi_t} \cdot d{\bf S}=
\int_{C_\Sigma} d^3p |\widehat{\Omega_-^\dagger\psi}({\bf p})|^2.
\end{equation}

The fundamental importance of the flux-across-surfaces theorem\ was first 
recognized by Combes,
Newton and Shtokhamer \cite{CNS}.  To our knowledge there exists no
rigorous proof of this theorem, although the heuristic argument for it is
straightforward. Let us consider first the
 ``free flux-across-surfaces theorem,'' where  $\psi_t := e^{-iH_0t}\psi$: 
\begin{equation}
\label{FASFREET}
\lim_{R \to \infty} \int_0^\infty dt \int_{R\Sigma} {\bf j}^{\psi_t} \cdot d{\bf S}=
\int_{C_\Sigma} d^3p |\hat{\psi}({\bf p})|^2
\end{equation}
(This free theorem, by the way, should be physically sufficient, since the 
scattered wave packet 
should in any case move  almost freely 
after the scattering has essentially been completed (see also \cite{CNS}).)

Now the current should contribute to the integral in (\ref{FASFREET}) only for large times,
because the packet must travel a long time before it reaches the distant
sphere at radius $R$. Thus we may use the long-time asymptotics of the free evolution.
We split $\psi_t({\bf x}) = (e^{-iH_0t} \psi)({\bf x})$ into
\begin{eqnarray}
\psi_t({\bf x}) &=& (\frac{m}{2\pi it})^{3/2}
\int d^3y   e^{im\frac{|{\bf x}-{\bf y}|^2}{2t}} \psi({\bf y})
\nonumber \\
&=& (\frac{m}{it})^{3/2}  e^{im\frac{x^2}{2t}}  \hat{\psi}(\frac{m{\bf x}}{t}) 
\nonumber \\ 
&+&  (\frac{m}{it})^{3/2}  e^{im\frac{x^2}{2t}} \int 
\frac{d^3y}{(2\pi)^{3/2}}e^{-im\frac{{\bf x}\cdot{\bf y}}{t}} 
(e^{im\frac{y^2}{2t}} - 1) \psi({\bf y}). \nonumber
\end{eqnarray}
Since
  $(e^{im\frac{y^2}{2t}} - 1)\to 0$ as $t\to\infty$,   we may neglect the second term,
so that as $t \to\infty$ we have that
\begin{equation}
\label{ASYMP}
\psi_t({\bf x}) \sim (\frac{m}{it})^{3/2} e^{im\frac{x^2}{2t}} \hat{\psi}(\frac{m{\bf x}}{t}).
\end{equation}
(This asymptotics  
has  long been recognized  as important for scattering theory, e.g.\
\cite{Brenig,Dollard}.)
From
(\ref{ASYMP}) we now find that
\begin{equation}
\label{JASYMPHEU}
{\bf j}^{\psi_t}({\bf x}) = \frac{1}{m} {\mbox{\rm Im}}  \/ \psi_t^*({\bf x}) 
\nabla \psi_t({\bf x}) \approx
\frac{{\bf x}}{t} (\frac{m}{t})^{3} |\hat{\psi}(\frac{m{\bf x}}{t})|^2. 
\end{equation}
(Note that by (\ref{JASYMPHEU}) the current
is strictly radial for large times, so that ${\bf j}^{\psi_t} \cdot d{\bf S}$
is indeed positive.)

Using now the approximation
(\ref{JASYMPHEU})  and substituting ${\bf p} := m\frac{{\bf x}}{t}$ we readily arrive 
at  
$$
\label{NICE}
\int_0^\infty dt \int_{R\Sigma} {\bf j}^{\psi_t} \cdot  d{\bf S} 
\approx 
\int_0^\infty dt \int_{R\Sigma} (\frac{m}{t})^3 |\hat{\psi}(\frac{m{\bf x}}{t})|^2  
\frac{{\bf x}}{t} \cdot  d{\bf S} $$ 

$$= \int_0^\infty dp p^2 \int_\Sigma d\sigma |\hat{\psi}({\bf p})|^2  
= \int_{C_\Sigma} d^3p |\hat{\psi}({\bf p})|^2.
$$

This heuristic argument for the free flux-across-surfaces theorem\ (\ref{FASFREET})
is so simple and intuitive that one may wonder
why  it does not appear in any primer on scattering theory. 
(For a rigorous proof see \cite{DDGZ1999}).

To arrive at the  general result  (\ref{FAST}) one may use the fact that 
the long time
behavior of $\psi_t({\bf x}) := e^{-iHt}\psi({\bf x})$ is governed by
$e^{-iH_0t}\Omega_-^\dagger \psi$ (see, e.g., \cite{Dollard})
so that the asymptotic current is simply
$$
{\bf j}^{\psi_t}({\bf x}) = {\mbox{\rm Im}}  \/ \psi_t^*({\bf x}) \nabla \psi_t({\bf x}) \approx
\frac{{\bf x}}{t} (\frac{m}{t})^3 |\widehat{\Omega_-^\dagger\psi}(\frac{m{\bf x}}{t})|^2,
$$
yielding (\ref{FAST}).

\section{Near Field Scattering} 

We turn now to a much more subtle question (see also \cite{leuven}): What
happens if we place the detectors {\it not} too distant  from the scattering
center and prepare the wave function near the scattering center, i.e., what
happens if we do not  take the limit $R\to\infty$ so central to scattering
theory? The detectors will of course again  fire at some random time and 
position, but what now of  the statistics?  This question is not quite
as innocent as it sounds; it concerns  in fact one of the most debated
problems in quantum theory: what we are considering here is the problem of
time measurement, specifically the problem of escape time (and position
at such time) of a particle from a region $G$. 
It is well known  that there is no
self-adjoint time observable of any sort and there is a huge and
controversial literature on this and on what to do about it. (See 
\cite{Busch,Buttiger}
and references therein.) Note also that since the exit position is the position of the
particle at a random time, it cannot be expressed as a Heisenberg
position operator in any obvious way.

The obvious answer (see \cite{Leavens} for a one-dimensional version) is,
of course, provided by (\ref{JDOTDSIGMA}), provided that the boundary of
$G$ is crossed at most once by the particle (whatever this is supposed to
mean for a quantum particle), so that every crossing of the boundary of $G$
is a first crossing, and provided of course that (\ref{JDOTDSIGMA}) is
nonnegative.\footnote{The wave function $\psi_t$ in (\ref{JDOTDSIGMA})
should of course be understood as referring to the Schr\"odinger evolution
with no detectors present.} Notice that the preceding provisos might well
be expected to be intimately connected. We thus propose that
(\ref{JDOTDSIGMA}) indeed gives the first exit statistics whenever the
following current positivity condition (a condition on both the wave
function and on the surface)
\begin{eqnarray}
{\rm CPC:} \quad
&&\forall t > 0 \quad \hbox{and} \quad \forall {\bf x}\in  \hbox{boundary
of the region \ } G \nonumber \\  &&{\bf j}^{\psi_t}({\bf x},t)  \cdot d{\bf S}
 > 0 \nonumber
\end{eqnarray}
is satisfied.

We predict that the statistics given by (\ref{JDOTDSIGMA}) will
(approximately) be obtained in an experiment on an ensemble of particles
prepared with (approximately) CPC wave function $\psi$ which is initially
well localized in some region $G$ whenever the detectors around the
boundary of $G$ (see Fig.~2) are {\it sufficiently passive\/}, a condition that needs to
be more carefully delineated but which should widely be satisfied. As to
how widely the CPC is satisfied, this is not easy to say. We do note,
however, that since whether or not it is satisfied depends upon the region
$G$ upon which we focus and around which we place our detectors, it may
often be possible to suitably adjust the region $G$ so that the CPC becomes
satisfied, at least approximately, even if the CPC fails to be satisfied
for our original choice of $G$.

A simple example of a situation where the CPC does hold and where one may
easily compute the exit-time statistics is the following. A  spherically
symmetric Gaussian wave packet, with initial width $\sigma$,  which is
initially located  at the center of $G$, a sphere with radius $R$,  evolves
freely. One  readily finds for the exit time probability density $\rho(t) :=
\int {\bf j}^{\psi_t} \cdot d{\bf S}$ that
$$
\rho(t) \propto \frac{R^3t}{\sigma^2} 
(\sigma^2 + (\frac{t}{2m\sigma^2})^2)^{-5/2} e^{-\frac{1}{2\sigma^2} 
\frac{R^2}{1+(\frac{t}{2m\sigma^2})^2}}. 
$$

Of course, some important questions remain: The expression
(\ref{JDOTDSIGMA}) is not a probability  for all wave functions---so what if
anything does it physically represent in general? And what in the general
case is the formula for the first exit statistics?

We stress again that the prediction (\ref{JDOTDSIGMA}) for the exit
statistics is not of the standard form, as given by the quantum formalism,
since it is not concerned with the measurement of an operator as
observable.\footnote{ Nor are they given by a positive-operator-valued
measure (POV), which has been proposed as a generalized quantum observable,
see \cite{Dav}} However, no claim is made that the expression
(\ref{JDOTDSIGMA}) and its interpretation cannot also be arrived at from
standard quantum mechanics---it presumably can---e.g., by including the
measurement devices in the quantum mechanical analysis. (See however
\cite{Busch}.)  After all, though there is no standard quantum observable
(i.e., self-adjoint operator) to directly describe the escape time, the
``pointer variable'' for the detectors {\it is\/} a standard quantum
observable, whose probability distribution after the experiment can in
principle be computed in the standard way.

In the next section we shall explain how the current as the central object
for escape and scattering phenomena arises naturally within Bohmian
mechanics \cite{Albert,DGZ}, where the physical meaning of
(\ref{JDOTDSIGMA}) turns out to be the measure for the expected number of
{\it signed\/} crossings, which of course can be negative.

\section{Bohmian mechanics}

In Bohmian mechanics   a particle moves along a trajectory ${{\bf x}}(t)$ determined by
\begin{equation}
\label{BM1}
\frac{d}{dt} {{\bf x}}(t) = {\bf v}^{\psi_t}({{\bf x}}(t)) =
\frac{1}{m} {\rm Im} \frac{{\bf \nabla} \psi_t}{\psi_t}({{\bf x}}(t)),
\end{equation}
where $\psi_t$ is  the particle's wave function, evolving according to
Schr\"odinger's equation. Moreover, if an ensemble of particles with wave function
$\psi$ is  prepared, the positions ${\bf x}$ of the particles are distributed
according to the quantum equilibrium measure 
$\mathchoice{ \hbox{${\rm I}\!{\rm P}$} }{ \hbox{${\rm I}\!{\rm P}$} }{ \hbox{$ 
\scriptstyle  {\rm I}\!{\rm P}$} }{ \hbox{$ \scriptscriptstyle  {\rm I}\!{\rm P}$} }^\psi$
  with density
$\rho=|\psi|^2$ ($\psi$ normalized) \cite{DGZ}.  

In particular, the continuity equation for the probability shows that the 
probability flux $(|\psi_t|^2,|\psi_t|^2 {\bf v}^{\psi_t})$
is conserved,
since 
$|\psi_t|^2 {\bf v}^{\psi_t} = {\bf j}^{\psi_t} $.

Hence, given $\psi_t$, the solutions ${{\bf x}}(t,{{\bf x}}_0)$ of equation
(\ref{BM1}) are  random trajectories, where the randomness comes from the
$\mathchoice{ \hbox{${\rm I}\!{\rm P}$} }{ \hbox{${\rm I}\!{\rm P}$} }{ \hbox{$ \scriptstyle  
{\rm I}\!{\rm P}$} }{ \hbox{$ \scriptscriptstyle  {\rm I}\!{\rm P}$} }   ^\psi$-distributed 
random initial  position ${{\bf x}}_0$, $\psi$ being the
initial wave function.

Consider now, at time t=0, a particle with wave function $\psi$ localized in some 
region $G\subset \mbox{${\rm I\!R}$}^3$ with
smooth boundary. 
Consider  the  number $N(dS,dt)$ of crossings by ${\bf x}(t)$ of the
surface element $dS$ of the boundary of $G$ in the
time $dt$ (see Fig.~3) .
Splitting $N(dS,dt) =: N_+(dS,dt) + N_-(dS,dt)$,
 where $N_+(dS,dt)$ denotes the number of outward crossings
and $N_-(dS,dt)$ the number of backward crossings of $dS$
in time $dt$,
we define  the number of signed crossings by
$N_s(dS,dt) =: N_+(dS,dt) - N_-(dS,dt).$

We can now compute the expectation values with respect to the
probability 
$\mathchoice{ \hbox{${\rm I}\!{\rm P}$} }{ \hbox{${\rm I}\!{\rm P}$} }{ \hbox{$ 
\scriptstyle  {\rm I}\!{\rm P}$} }{ \hbox{$ \scriptscriptstyle  
{\rm I}\!{\rm P}$} }^\psi$
of these numbers in the usual statistical mechanics manner.
Note that for a crossing of $dS$ in the  time interval $(t,t+dt)$ to occur, 
the particle has to be
in a cylinder of size $|{\bf v}^{\psi_t} dt  \cdot d{\bf S}| $ at time $t$. 
Thus we obtain for the expectation value
\begin{eqnarray}
\label{EXPECT}
\mathchoice{ \hbox{${\rm I}\!{\rm E}$} }{ \hbox{${\rm I}\!{\rm E}$} }{ \hbox{$ 
\scriptstyle  {\rm I}\!{\rm E}$} }{ \hbox{$ \scriptscriptstyle  
{\rm I}\!{\rm E}$} }^\psi(N(dS,dt)) =    |\psi_t|^2 |{\bf v}^{\psi_t} dt
\cdot d{\bf S}|      =  |{\bf j}^{\psi_t} \cdot
d{\bf S}|dt , \nonumber 
\end{eqnarray} 
and similarly 
$ \mathchoice{ \hbox{${\rm I}\!{\rm E}$} }{ \hbox{${\rm I}\!{\rm E}$} }{ \hbox{$ 
\scriptstyle  {\rm I}\!{\rm E}$} }{ \hbox{$ \scriptscriptstyle  
{\rm I}\!{\rm E}$} }^\psi(N_s(dS,dt)) =
 {\bf j}^{\psi_t} \cdot d{\bf S} dt. $

If we further introduce the random variables $t_e$, the  first exit time
>from $G$, $t_e := \inf \{t\ge 0| {\bf x}(t) \notin G\}$, 
and ${\bf x}_e$, the  position of first exit,  
${\bf x}_e= {\bf x}(t_e)$,
we obtain a very natural and principled explanation of what
we arrived at in a heuristic and suggestive manner in our treatment of
scattering theory and the statistics of the first exit time and position. 
For Bohmian mechanics the CPC implies that 
every trajectory crosses  the boundary of $G$  at most once, and in this  case
we have   
\begin{eqnarray}
\mathchoice{ \hbox{${\rm I}\!{\rm E}$} }{ \hbox{${\rm I}\!{\rm E}$} }{ 
\hbox{$ \scriptstyle  {\rm I}\!{\rm E}$} }{ \hbox{$ \scriptscriptstyle  
{\rm I}\!{\rm E}$} }^\psi(N(dS,dt) ) &=& \mathchoice{ \hbox{${\rm I}\!{\rm E}$} 
}{ \hbox{${\rm I}\!{\rm E}$} }{ \hbox{$ \scriptstyle  {\rm I}\!{\rm E}$} 
}{ \hbox{$ \scriptscriptstyle  {\rm I}\!{\rm E}$} }^\psi((N_s(dS,dt))=
\nonumber \\
 0\cdot\mathchoice{ \hbox{${\rm I}\!{\rm P}$} }{ \hbox{${\rm I}\!{\rm P}$} 
}{ \hbox{$ \scriptstyle  {\rm I}\!{\rm P}$} }{ \hbox{$ \scriptscriptstyle 
 {\rm I}\!{\rm P}$} }   ^\psi(t_e\notin dt {\mbox{\rm \ or }} {\bf x}_e\notin dS) 
&+& 1 \cdot \mathchoice{ \hbox{${\rm I}\!{\rm P}$} }{ \hbox{${\rm I}\!{\rm P}$} }{ 
\hbox{$ \scriptstyle  {\rm I}\!{\rm P}$} }{ \hbox{$ \scriptscriptstyle  
{\rm I}\!{\rm P}$} }   ^\psi({\bf x}_e\in dS {\mbox{\rm \ and }}
t_e\in dt) \nonumber
\end{eqnarray}
and we find  for the joint {\it probability} of exit
through  $dS$ in  time $dt$
\begin{equation}
\label{POPRAN}
\mathchoice{ \hbox{${\rm I}\!{\rm P}$} }{ \hbox{${\rm I}\!{\rm P}$} 
}{ \hbox{$ \scriptstyle  {\rm I}\!{\rm P}$} }{ \hbox{$ 
\scriptscriptstyle  {\rm I}\!{\rm P}$} }   ^\psi({\bf x}_e\in dS 
{\mbox{\rm \ and }} t_e\in dt) = 
 {\bf j}^{\psi_t}  \cdot d{\bf S} dt. 
\end{equation}

In principle one could compute the first exit statistics also when the CPC
fails to be satisfied.  These are in fact given by the same formula
(\ref{POPRAN}) as before, provided one replaces $ {\bf j}^{\psi_t}$ by the
truncated probability current $\tilde{\bf j}$ arising from killing the
particle when it  reaches the boundary of $G$.  This is simply given, on the
boundary of $G$, by
\begin{equation}
\label{shit}
 { \tilde{\bf j}}^{\psi_t}(t,{\bf x})= \left\{ \begin{array}{ccc} {\bf
j}^{\psi_t}({\bf x})&\quad & \mbox{if $(t,{\bf x})$ is a first exit from
$G$}\\ 0&\quad & \mbox{otherwise} \end{array} \right.
\end{equation}
 where $(t,{\bf x})$ is a first exit from $G$ if the Bohmian trajectory
passing through ${\bf x}$ at time $t$ leaves $G$ at this time, for the
first time since $t=0$.  Thus, we have generally that
\begin{equation}
\label{POPshit}
\mathchoice{ \hbox{${\rm I}\!{\rm P}$} }{ \hbox{${\rm I}\!{\rm P}$} 
}{ \hbox{$ \scriptstyle  {\rm I}\!{\rm P}$} }{ \hbox{$ 
\scriptscriptstyle  {\rm I}\!{\rm P}$} }   ^\psi({\bf x}_e\in dS 
{\mbox{\rm \ and }} t_e\in dt) = 
\tilde{ {\bf j}}^{\psi_t}  \cdot d{\bf S} dt. 
\end{equation}

However, there is an important difference between the CPC probability
formula (\ref{POPRAN}), involving the usual current, and the formula
(\ref{POPshit}), involving the truncated current.  The usual current is well
defined in orthodox quantum theory, even if it is true, as we argue, that
its full significance can only be appreciated from a Bohmian
perspective. The truncated current cannot even be {\it defined\/} without
reference to Bohmian mechanics, since whether or not $(t,{\bf x})$ is a
first exit from $G$ depends upon the full and detailed trajectory up to
time $t$. (In particular, a different choice of dynamics, as for example
given by stochastic mechanics \cite{Nel85,Gol87}, would yield a different truncated
current. It is natural to wonder whether the truncated current given by
Bohmian mechanics provides in the general case the best fit to the
measured escape statistics expressible without reference to the measuring
apparatus.)
 
\bigskip

Finally, we note that in the context of scattering theory our definition
(\ref{SIGMAFLUX}) of $\sigma_{{\mbox{\rm \footnotesize flux}}}$ captures
exactly what it should once one has real trajectories, namely the
asymptotic probability distribution of exit positions,
$$
\sigma_{{\mbox{\rm \footnotesize flux}}}(\Sigma) = 
\lim_{R \to \infty} \mathchoice{ \hbox{${\rm I}\!{\rm P}$} 
}{ \hbox{${\rm I}\!{\rm P}$} }{ \hbox{$ \scriptstyle  
{\rm I}\!{\rm P}$} }{ \hbox{$ \scriptscriptstyle  
{\rm I}\!{\rm P}$} }^\psi({\bf x}_e\in R\Sigma).  
$$
This follows from the fact that  the expected number of backward crossings
of the sphere of radius $R$ vanishes as $R\to\infty$ (see \cite{DDGZ1999}).


\begin{figure}
\label{STREUNG}
\begin{center}
\leavevmode%
\epsfxsize=10cm
\epsffile{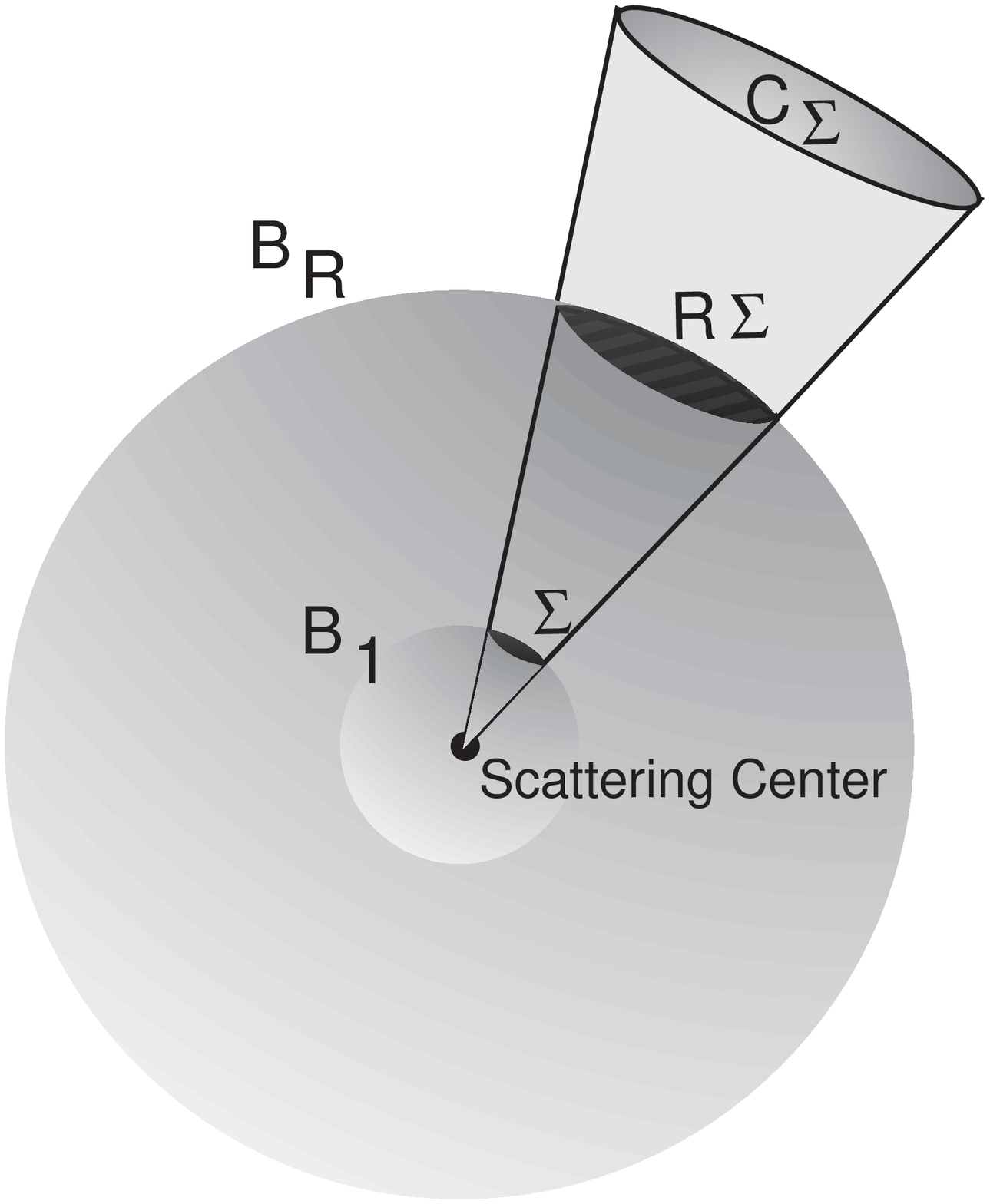}
\end{center}
\caption{The geometry of the scattering-into-cones and the flux-across-surfaces theorems.}
 \end{figure}

\newpage

\begin{figure}
\label{Bohm}
\begin{center}
\leavevmode%
\epsfxsize=10cm
\epsffile{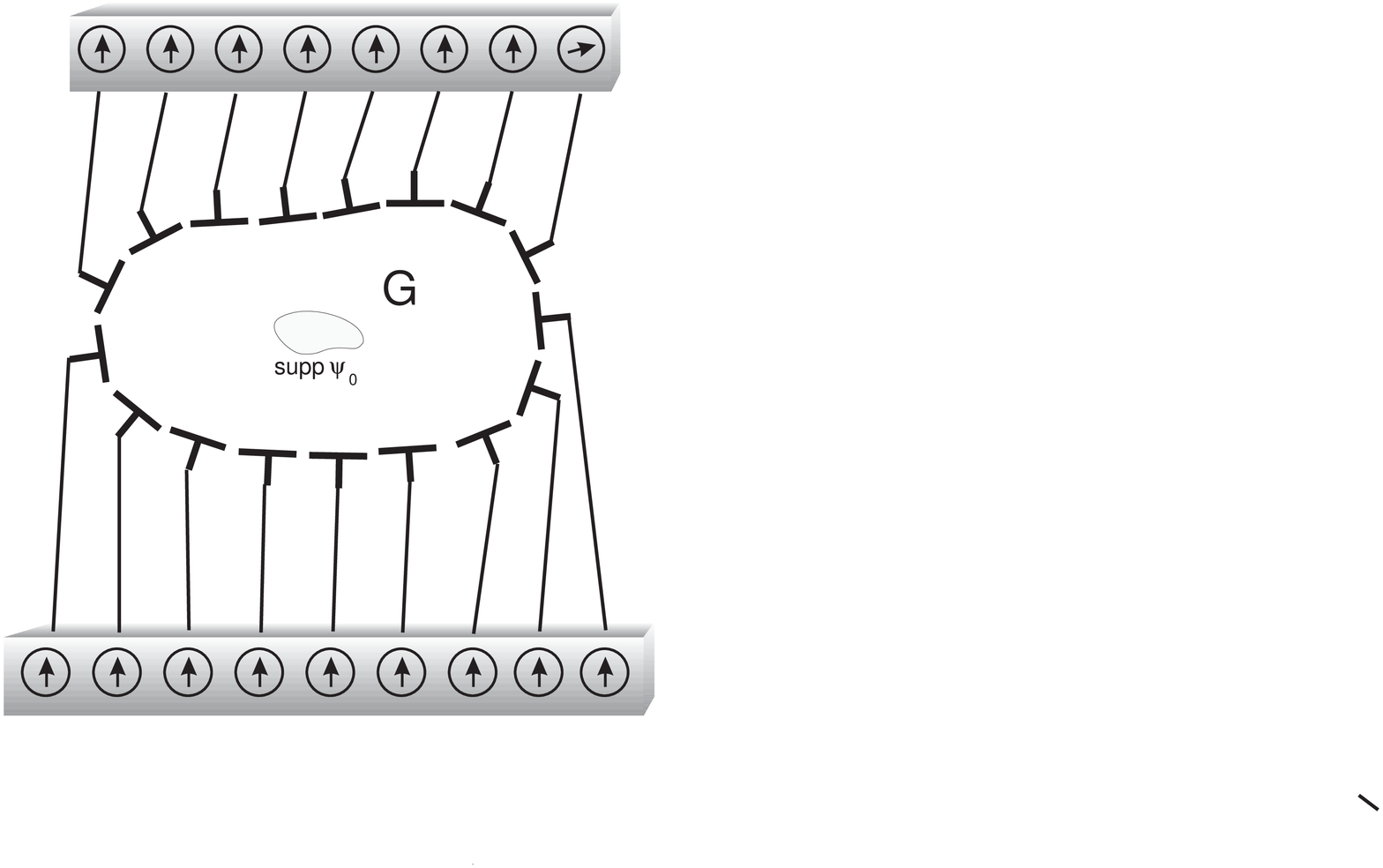}
\end{center}
\caption{Escape experiment: A region $G$ is defined by an array of
detectors, which surround a smaller region, supp\,$\psi_0$, in which a
particle's wave function is initially localized. The detectors record the
time at which they fire. Typically only one of the detectors will fire, and
the position of this detector yields the measured exit position.}
\end{figure}

\newpage

\begin{figure}
\label{Bohm1}
\begin{center}
\leavevmode%
\epsfxsize=10cm
\epsffile{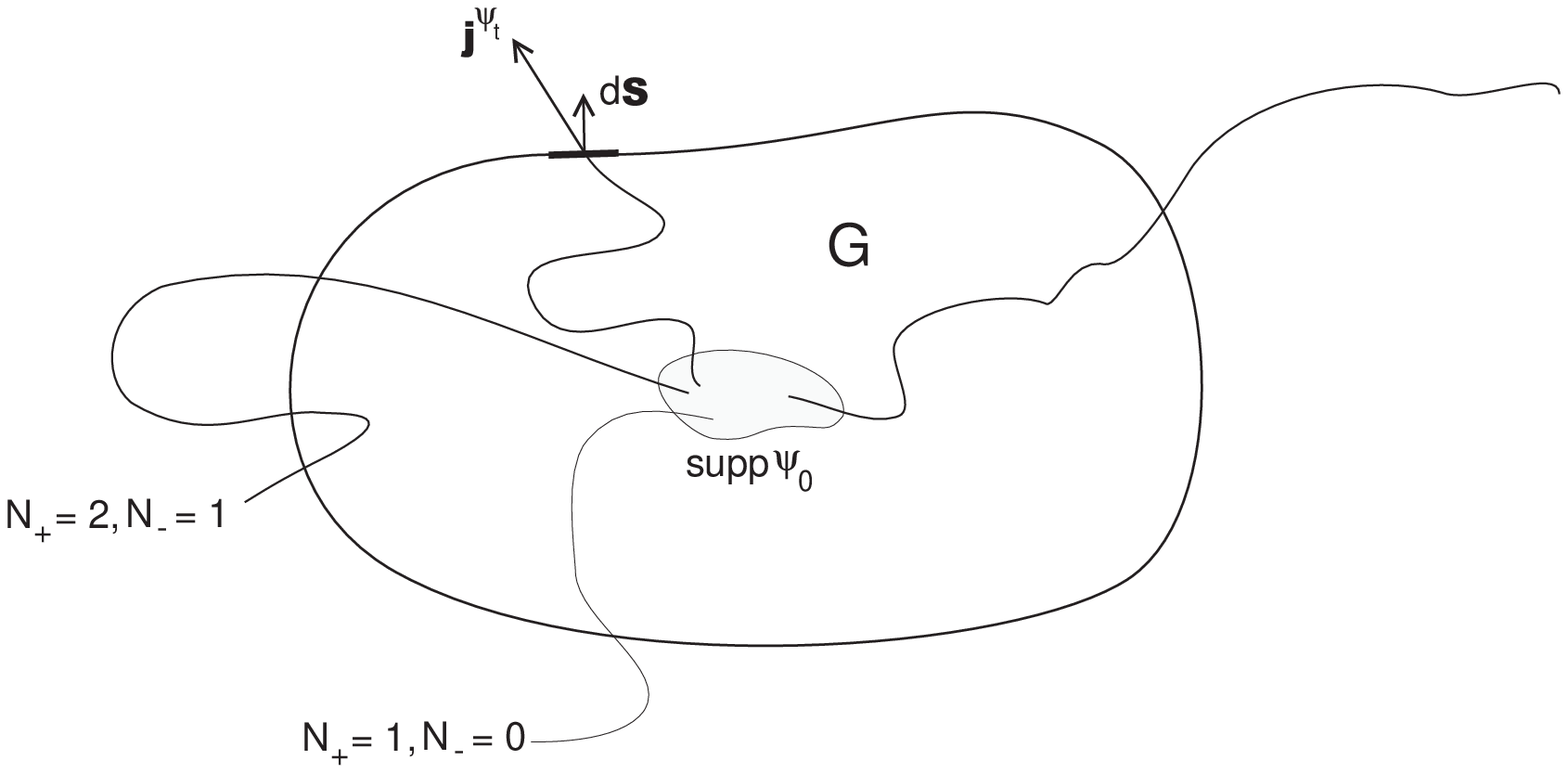}
\end{center}
\caption{In Bohmian mechanics the flow lines of the current represent the
possible trajectories of the Bohmian particle. Some Bohmian trajectories
leaving $G$ are drawn (for the Schr\"odinger evolution without detectors,
see footnote 1).}
\end{figure}

\end{document}